# Innovation ARIMA models application to predict pressure variations in water supply networks with open-loop control. Case study in Noja (Cantabria, Spain)

David Muñoz-Rodríguez [a], Manuel J. González-Ortega [b], María-Jesús Aguilera-Ureña [a], Andrés Ortega-Ballesteros [a], Alberto-Jesus Perea-Moreno a a Departamento de Física Aplicada, Radiología y Medicina Física, Universidad de Córdoba, Campus de Rabanales, 14071 Córdoba, Spain

b Department of Aerospace Engineering and Fluid Mechanics, University of Seville, 41013 Seville, Spain

**Abstract:** Water utilities are increasingly concerned about losses, leaks, and illegal connections in their distribution networks. Pressure control is typically managed through pressure reducing valves (PRVs) with electrically controlled actuators based on predefined tables according to the pressure at the critical point control (CPC). This open-loop control method lacks direct feedback between the PRV and CPC, making it challenging to distinguish whether pressure variations originate from normal head losses or abnormal network conditions.

Unlike traditional applications of ARIMA focused on water demand forecasting, this study explores its novel use in pressure management within distribution networks, aiming to predict P3 (CPC) pressure based on head losses across a defined hydraulic sector. To achieve this objective, a predictive model based on the Box-Jenkins methodology and its variations is implemented to analyse time series data. An action path is established to determine the optimal model—ARIMA, ARMA, ARMAX, etc.—which is subsequently validated using real operational data from Noja, a coastal town in northern Spain characterized by significant seasonal population fluctuations. By accurately forecasting CPC pressure, this system enhances the detection of anomalous patterns, enabling more efficient network pressure management. The study demonstrates the potential of advanced modelling techniques in optimizing water distribution networks, providing valuable insights to improve system efficiency, reliability, and sustainability in urban environments.

## 1. Introduction

The proper and right management of water resources is gaining more and more importance over the last years since it is required to secure drinking water supply to the growing world population, which, in many cases, is concentrated in water scarce areas or places with less water available [1-4]. One of the main pillars to secure water supply management is leakage control to avoid water losses [5-7].

The International Water Association (IWA) differentiates four key points in a leakage policy management strategy: active leakage control, asset management, speed and quality of repairs and pressure management [8]. Pressure management has proven to be the most effective method for reducing and preventing the appearance of new leaks in drinking water distribution piping systems due to the direct relationship between the working pressure in the network and the water lost through leaks, as well as the appearance of new leaks [9,10]. Thus, controlling the pressure in the networks allows to manage leakage level [11,12]. The relationship between the ratio of pressure reduction in the network and leakage decrease depends on several factors, such as pipe material and the pressure reduction level [13,14]. Moreover, pressure management is also employed to manage customer demand an ensure adequate water supply throughout the day, though this application is not so widely extended [15,16].

The most common way to apply pressure management is a water distribution network is by the use of pressure reducing valves (PRV) [17,18,4,19]. The function of a pressure reducing valve is to maintain a fixed downstream pressure independent of flow variations or upstream pressure oscillations occurring in the network. Over the last years, electronic control systems and algorithms, with different levels of complexity, have been introduced and developed to optimize the performance of these valves [20]. In this respect, it must be added that recently pumps as turbines (PAT) have also become an optimal and sustainable alternative to PRVs [21] since they also meet energy saving targets. And other authors such as Fontana and Marini (2021) propose an approach to integrating hydropower generation with pressure-reducing valves in water distribution networks. Instead of dissipating excess head through PRVs, this approach generates power. Utilising a programmable logic controller (PLC), they address the optimisation problem directly, circumventing the necessity for external modules [22].

These pressure control valves have different strategies for pressure management: fixed outlet pressure, time-based, flow-based, and critical point control (CPC). Of all of them, CPC is the most advanced, the one that achieves a control more adjusted to the dynamics of the network, but it is also the most complex [23,24]. The aim of CPC systems is keeping the pressure in the area as low as possible to reduce leakages, while assuring the right level of supply to customers. Under CPC, the PRV is adjusted dynamically with different values at the outlet to maintain a fixed pressure at the critical point of the area supplied [25]. CPC systems can be classified into open or closed loop, depending on whether there is real time connection between the critical point and the pressure reducing valve or not [26,27,4]. In critical point close loop, the critical point and the PRV are communicating continuously to set the pressure value at the outlet of the valve according to the pressure at the critical point. This type of systems requires infrastructure for real time communication between the critical point and the PRV, which, in most of the cases, is expensive and difficult to commission. In open-loop systems, the pressure reducing valve and the critical point do not share real time signals and are based on using statistical models and time series analysis, to forecast hydraulic performance of the network.

In open-loop systems, the pressure reducing valve and the critical point do not share real-time signals and are based on using statistical models and time series analysis to forecast the hydraulic performance of the network. Several studies have explored the use of statistical techniques to detect and classify anomalies in water distribution systems, distinguishing between events such as leaks, pipe bursts, and unauthorized water usage. For instance, [55] proposed a method based on transform analysis in wireless sensor networks to detect pipeline bursts and leaks, while [56] reviewed pressure-based techniques for leak monitoring in pipe distribution systems. These studies focus on identifying the causes of anomalies in water supply networks, which differs from our approach. In contrast, this study applies ARIMA models not for anomaly classification but for pressure management by predicting P3 pressure based on head losses, optimizing the efficiency of pressure control strategies.

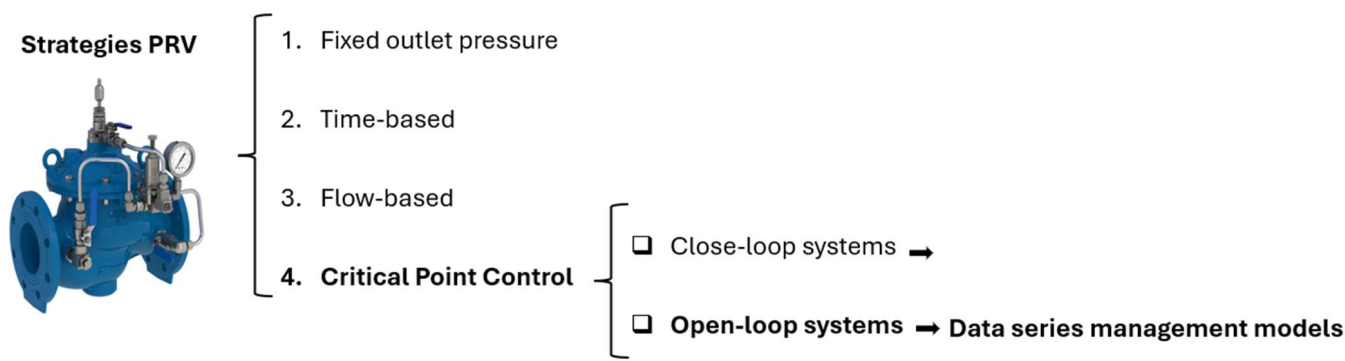

**Figure 1.** Strategies for Pressure Reducing Valves (PRV) in Water Distribution Systems.

There are different references that shows that the Box and Jenkins methodology (1976) is very suitable for the analysis of the data service to predict the hydraulic performance of the network [28,29], in PRV with open loop control strategy. This methodology proposed by Box and Jenkins is a statistical method used to analyse and predict time series and is part of the general ARMA model. Their work, developed in the first five years of the 1970s (Figure 2), has led to one of the methodologies most widely used in the predictive analysis of numerical series [30]. The methodology has essentially three stages. Stage 1 consists of identifying the model type and for this it is important to detect whether the data present patterns, trends, seasonal behaviour or whether the series is stationary or not. Stage 2 consists of identifying the coefficients or parameters of the model using maximum likelihood techniques and the last one consists of evaluating the accuracy of the model.

The general model from which the Box and Jenkins methodology originates is composed of two models, the AR (Autoregressive) model and the MA (Moving Average) model. The AR term represents the relationship between a current observation and previous observations, and the MA term indicates the relationship between a current observation and the residual error of the moving average of previous observations. Each of these two terms is represented by a numerical value that represents the number of past values or errors of past values that influence the current value. The AR term is assigned the letter *p* and the MA is assigned the letter *q*.

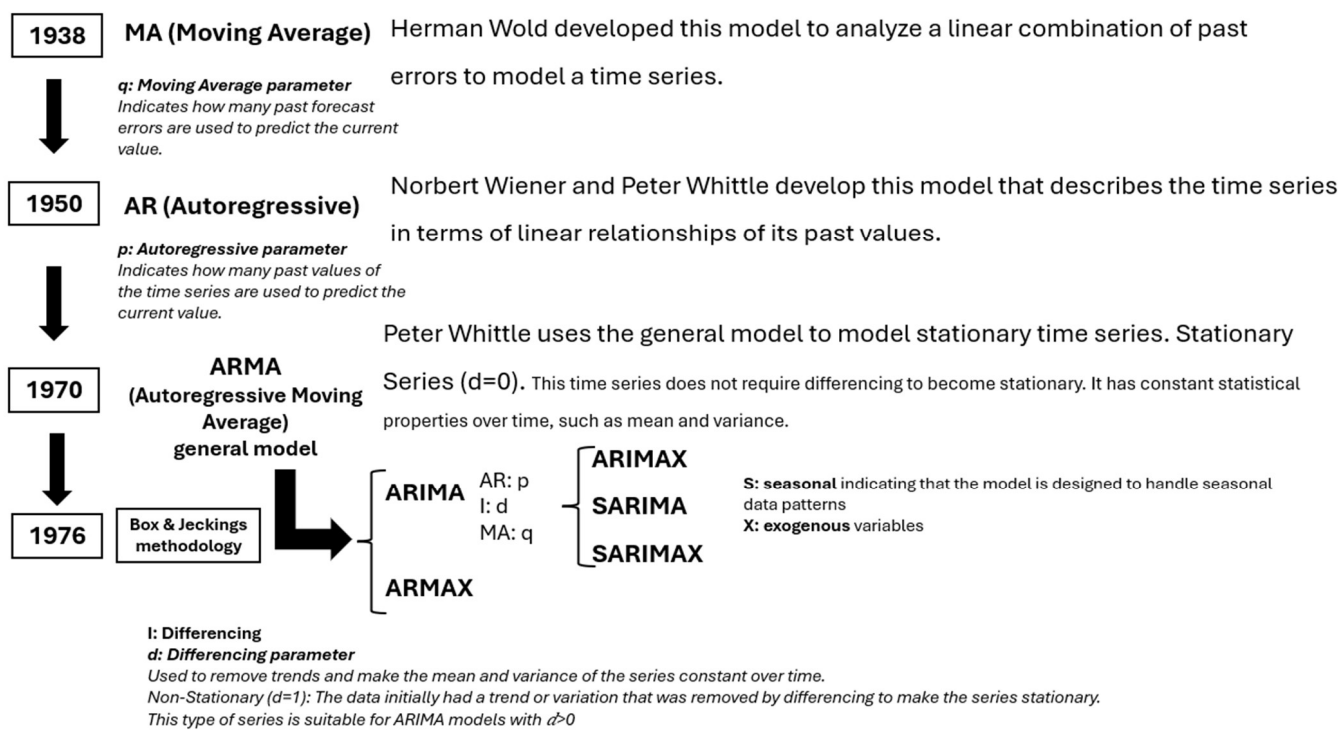

**Figure 2.** Evolution of Time Series Models: AR, MA, and ARMA.

The above general model presents variants depending on whether the time series considered in the modelling is stationary or not, whether it presents seasonal behaviour and/or whether there are exogenous factors that may influence the variable studied. According to the above, the existing variables are called ARIMA, SARIMA, ARIMAX, ARMAX or SARIMAX. The I present in any of the models refers to the differentiation operation that may be carried out on the series. The seasonal component of the series is represented by the letter S in the name of the model. If the series under study presents this behaviour, the model will be assigned three coefficients (P, D, Q) representing the relationship between present and past observations (P) and the relationship with the residual errors of paired values (Q). The (D) refers to the level of seasonal differentiation made in the original series. If, in addition to the above, the models present an X at the end, it means that they present exogenous variables [30].

There are different uses for the general ARMA model and its ARIMA variant, with or without exogenous variables. The studies performed with ARIMA models and their adaptations are diverse in different research fields. Zhou et al. [31], Wong et al. [32] and Du et al. [33] used these models to forecast water consumption. Liu et al. [34] studied acute haemorrhagic conjunctivitis and Makoni et al. [35] modelled tourism inflow in Zimbabwe. Ampountolas [36] modelled and forecasted daily customer demand in a hotel using, among others, a SARIMAX model. Manigandan et al. [37] forecasted gas consumption in the United States. Yin et al. [38] use an ARMAX model for output noise estimation in control engineering. Lim et al. [39] consider the ARMAX model to estimate the travel demand of Japanese citizens to destinations such as New Zealand and Taiwan.

The present study conducts an analysis of the data series from the Noja (Cantabria, Spain) potable water network. With more than 80,000 real data points from the network, the best model is fitted and validated using the Box and Jenkins methodology. The objective is to apply an innovative development of time series analysis to open-loop critical point pressure management system to predict the pressure at the critical point (P3). The pressure at the critical point (P3) depends on two exogenous variables: pressure at the outlet of the pressure-reducing valve (P2), which is adjusted by the pressure reducing valve (PRV) and the flow rate at the inlet of the area (Q), which depends of the total consumption demanded by the population. The model will employ these two variables, P2 and Q to forecast the pressure at the critical point (P3). The model will be applied with real data of a hydraulic network with seasonal demand where critical point pressure control is essential to deliver water with the right level of service while minimizing leakages rates.

## 2. Materials and Methods

*2.1. Description of Case Study area*

This study's series have been analysed and validated with real data provided by the Noja water company. Noja (Figure 3) is a coastal town in the Cantabria region, northern Spain. This town has a permanent population of approximately 2700 inhabitants, which increases by more than 3600 % during the summer, reaching up to 100,000 people during the holiday period. Therefore, Noja has a very seasonal demand with great variation between winter and holiday periods, such as Easter or summer.

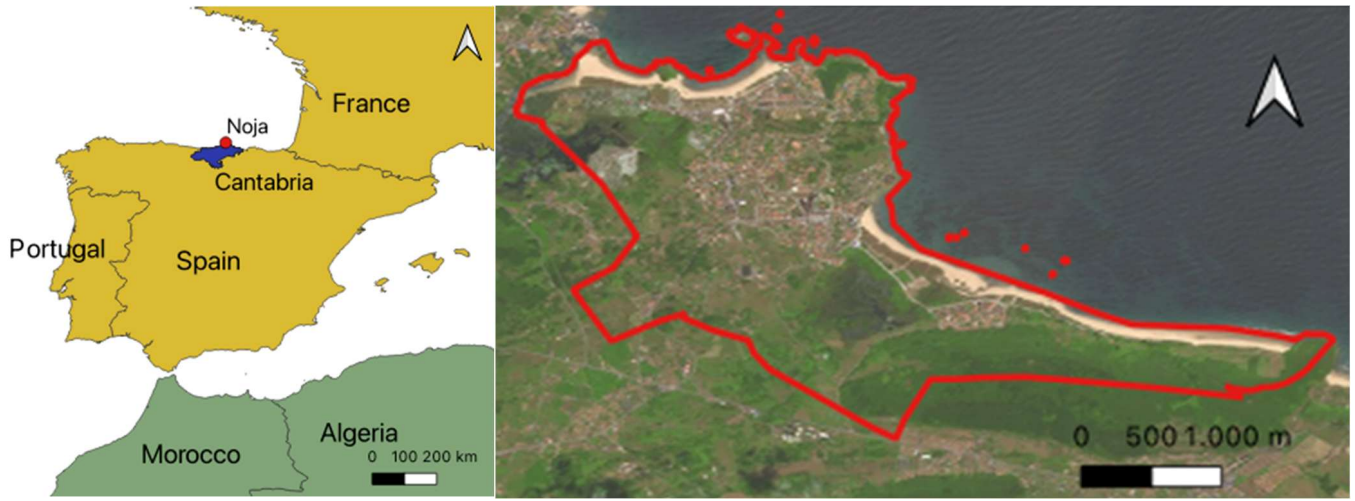

**Figure 3.** Location of the study area (left), municipality on orthophoto (right).

*2.2. Description of the hydraulic control of the water supply in Noja*

The water supply system in the town of Noja operates entirely by gravity from a distribution tank. The total length of the water distribution network (WDN) is around 50 kilometres. The main pipe materials are predominantly ductile iron (DI) and polyethylene (PE), accounting for approximately 90% of the total length. Asbestos (AB) and polyvinyl chloride (PVC) pipes make up the remaining 10%. Larger sizes, from DN150 to DN300, are constructed of metallic materials, while plastic materials are used for sizes below DN150.

Figure 4 shows the variations in flow ($m^3/h$) demanded from the 28 February 2021 to 20 August 2022, measured every 15 minutes. The measurements come from an electromagnetic flowmeter installed at the inlet of the water distribution system of the town. The technical specification of the flowmeter is displayed in table 2. It can be seen how in July and August the flow supplied increases up to 130 $m^3/h$. However, during the month of March the maximum flow is 20 $m^3/h$. Between the first and second quarter of the year 2022, there is also an increase in water supplied due to the Easter week, a holiday period in Spain.

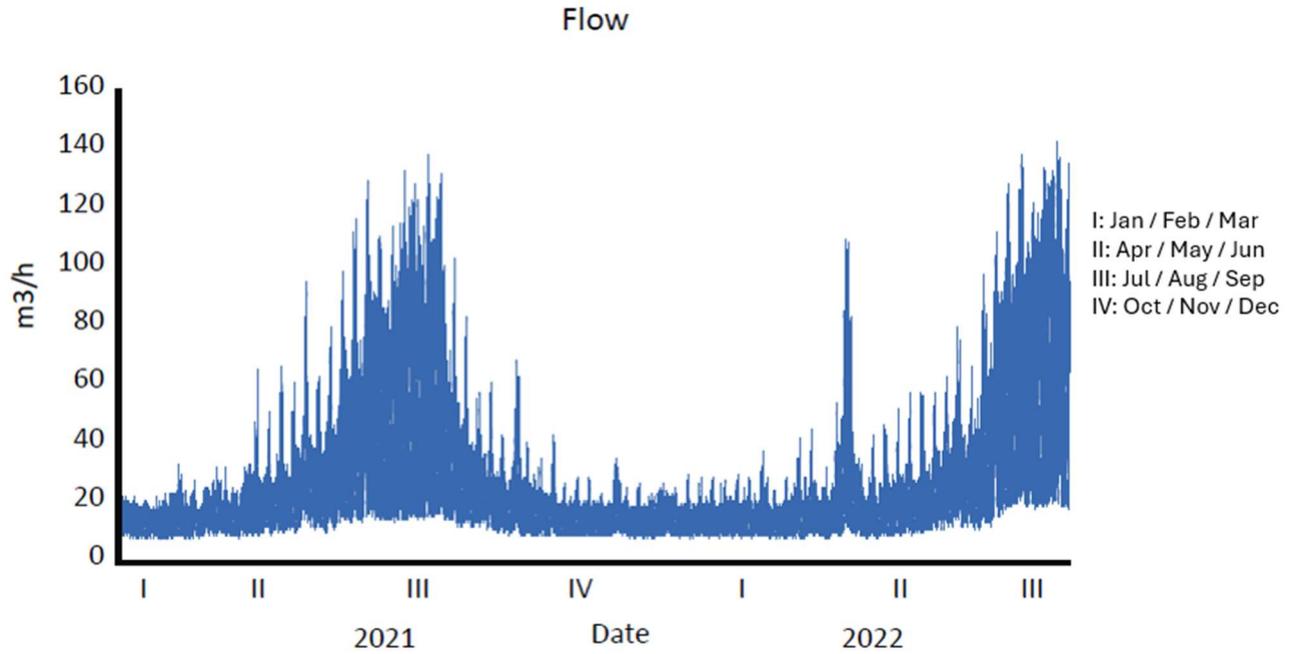

**Figure 4.** Flow variations from March 2022 to October 2022.

The distribution tank supplying the water system is located at the highest point of the municipality, creating a geometric difference that results in an initial pressure of approximately 60 m (P1). This pressure, known as the upstream pressure of the pressure-reducing valve, is generated before the valve reduces the pressure required for the system. Therefore, P1 is the pressure upstream of the pressure-reducing valve, which must be controlled. The water supply is regulated by a DN300 (12") full-bore, membrane-actuated hydraulic control valve. The hydraulic specifications of this valve are detailed in Table 1.

**Table 1.** Hydraulic specification of the valve.

| Features | Specification |
|---|---|
| Valve type | Globe, straight flow, diaphragm- actuated, Resilient disc seal, Flange connections |
| Control function | Pressure reducing |
| Size | 12" (300 mm) |
| Max Flow rate | 1590 m$^3$/h |
| Min Flow rate | 11,50 m$^3$/h |
| Weight | 528 Kg |
| Fluid type | Potable water |
| Connection | ISO PN16 |

**Table 2.** Electromagnetic flowmeter specification.

| Features | Specification |
|---|---|
| Flowmeter type | Electromagnetic |
| Size | 12" (300 mm) |
| Accuracy | ±0.4% ±2 mm/s |
| Measurement | Bidirectional |
| Weight | 88 Kg |
| Fluid type | Potable water |
| Connection | ISO PN16 |

The control loop of the valve includes two systems in parallel: a standard mechanical pressure reducing pilot and external electronic controller to command the pressure at the outlet of the valve, at point P2 (42 m), shown in Figure 5, in order to maintain a target pressure at the critical point P3 (25 m), according to the values in Table 3, depending on the time of day and day of the week. During normal operation, the pressure reducing pilot is isolated from the control chamber of the valve, so the electronic controller is working and controlling the pressure at the outlet of the valve. The mechanical pressure reducing pilot is used as a backup is case of failure or maintenance with the electronic controller.

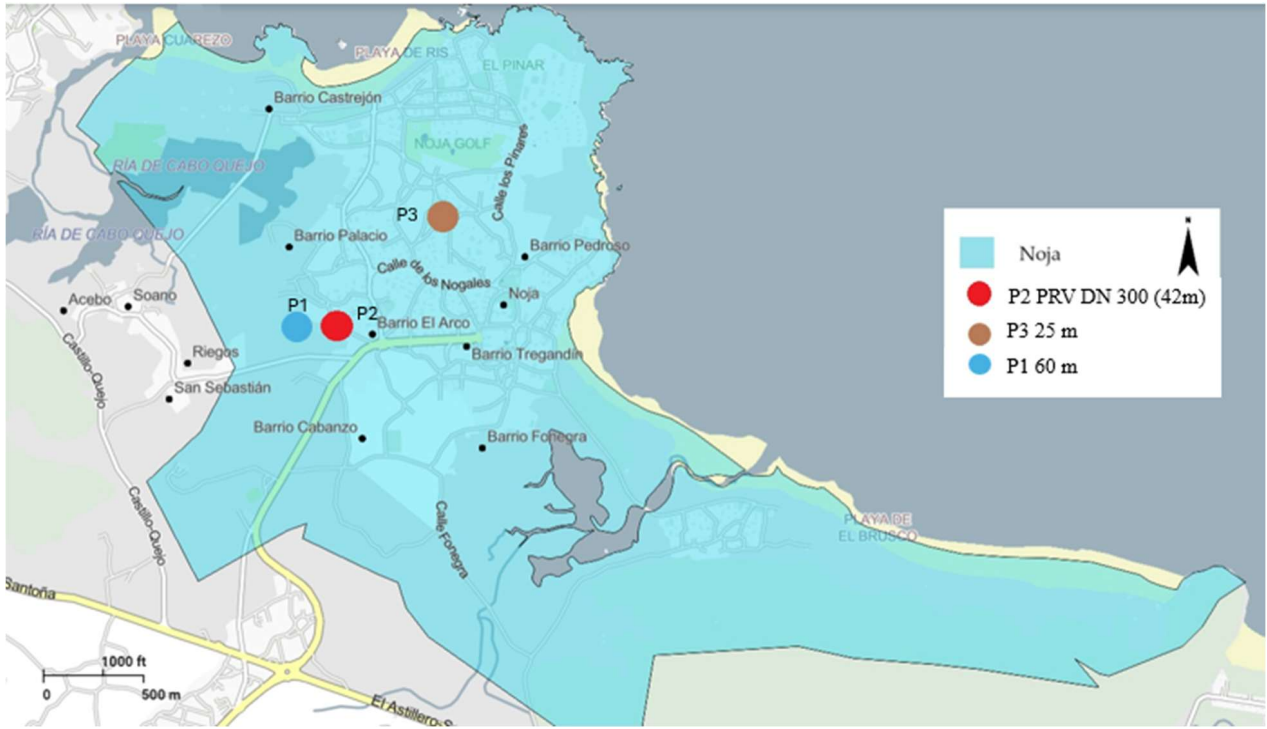

**Figure 5.** P3 and PRV location in Noja and the area supplied by the PRV.

The P3 target critical points values are selected by the water operator. The P3 values are chosen to secure the right level of service in the area. During the central time of the day, when the demand is higher, the P3 values are higher to secure optimal service to the customers. During the night period, when the consumption is lower and demand reduced, the P3 values are lower to minimize water lost by leakages. This PRV is based on an open-loop Critical Point Control (CPC) method, where pressure modulation at P2 depends on flow variations to ensure pressure stabilisation at P3 (Figure 6). The flow data essential for this control mechanism is derived from the pulse output of a water meter located downstream of the PRV. To prevent interference with the valve's original pressure reduction pilot, the electronic controller is installed in parallel and isolated. In addition, upstream and downstream pressures are recorded at 15 minutes intervals with a sampling frequency of 10 seconds. The controller adjusts the valve outlet pressure according to the prescribed pressure management protocol, maintaining an industry standard deadband of ±0.5 m. A data logger positioned at the critical point records pressure data using the same logging interval and sampling rate.

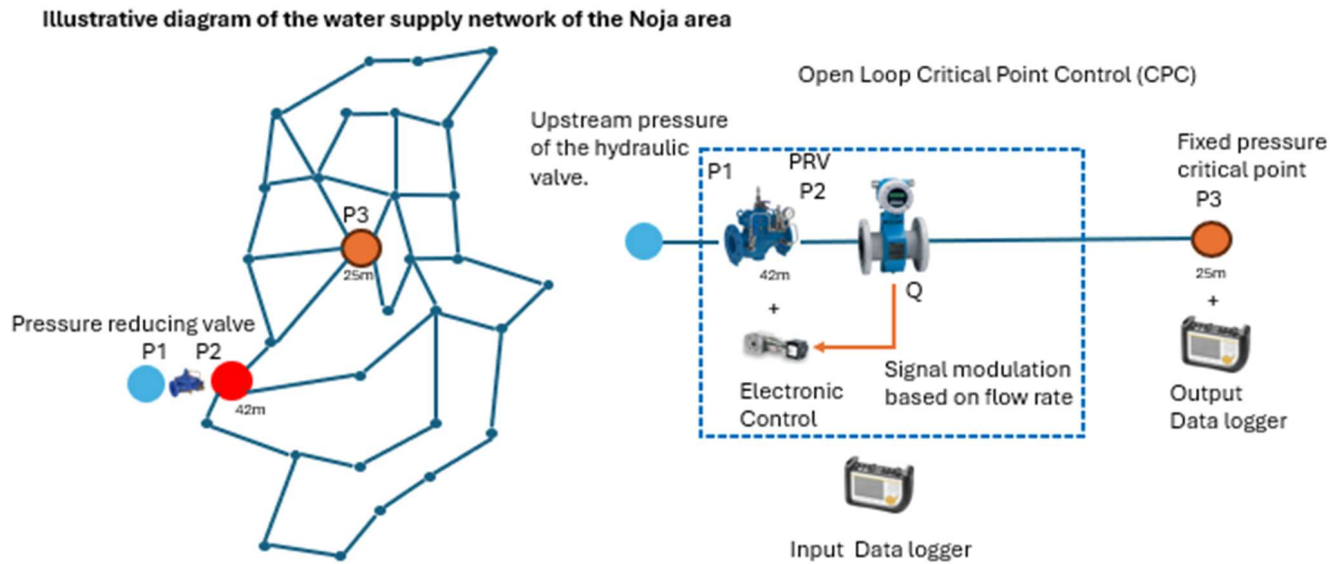

**Figure 6.** Open-loop Critical Point Control.

The water company knows the value that P3 should have based on the flow demands, which vary throughout the day and also depend on the time of year, with an increase in population during the summer period. For example, at night when the water demand decreases (Q, flow down), the pressure in the network significantly increases. During the day, at peak demand times, as the flow rate increases, the pressure P3 decreases, which requires regulation at P2 to maintain a minimum service pressure. These pressure patterns are known by the experience of the water operator and can be used to determine if there are leaks in the network, provided a model can be established that fits these patterns from the time series analysis. In Table 3, the target pressure that the water company must maintain can be seen. As displayed in the table, during peak demand hours, it must maintain a high pressure to sustain the service (flow rate and pressure). As shown in Table 3 below, the operator selected the critical point target pressure based on their familiarity with the system and the surrounding conditions. The operator distinguishes between working days and weekends by maintaining distinct P3 target values based on the time of day.

**Table 3.** Critical pressure values at P3 (m).

| Hours | Monday to Friday | Weekend |
| --- | --- | --- |
| 00:00-01:00 | 28 | 28 |
| 01:00-02:00 | 28 | 22 |
| 02:00-05:00 | 22 | 22 |
| 05:00-06:00 | 24 | 23 |
| 06:00-07:00 | 24 | 26 |
| 07:00-08:00 | 32 | 26 |
| 08:00-09:00 | 32 | 33 |
| 09:00-12:00 | 33 | 33 |
| 12:00-13:00 | 32 | 33 |
| 13:00-14:00 | 32 | 32 |
| 14:00-00:00 | 33 | 32 |

Figure 7 displays hydraulic performance of the network during the first two weeks of March 2021. It can be seen how the pressure at the outlet of the pressure reducing valve, P2, changes dynamically according to the flow to keep the target pressure defined by the water utility at the critical point, P3.

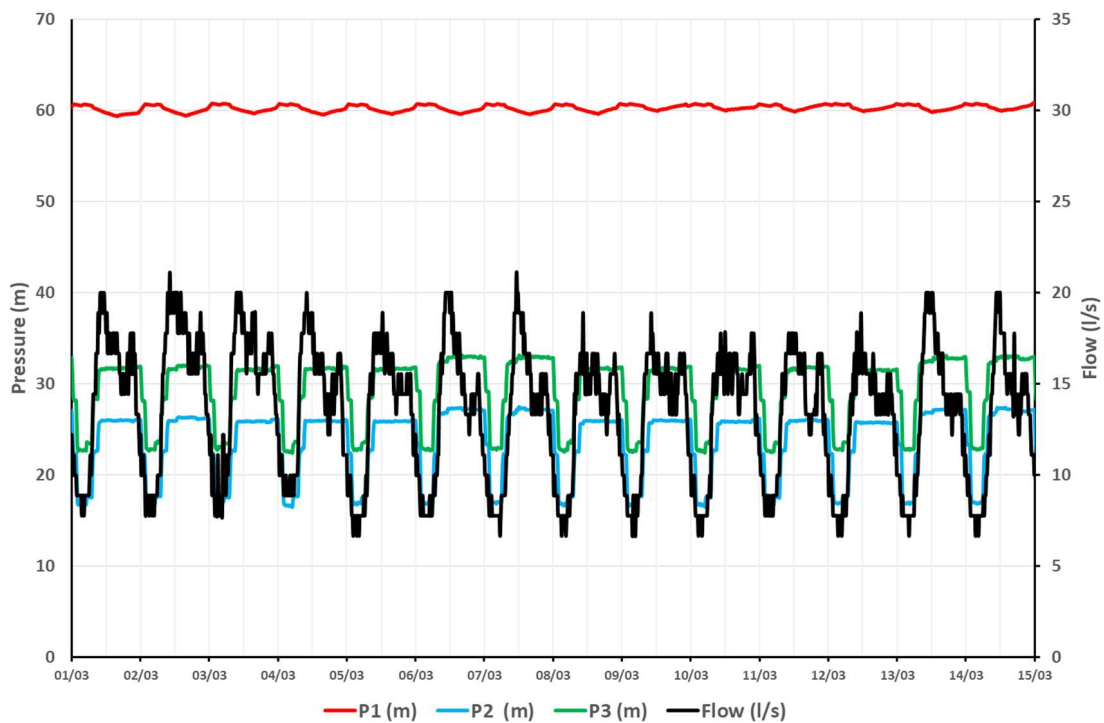

**Figure 7.** Hydraulic performance of the network during the first two weeks of March 2021.

The aim of this study is to develop and evaluate an innovative system for pressure management at the critical point P3 of a drinking water distribution network using the Box Jenkins methodology and the model that best fits the characteristics of the time series, ARMA or ARMAX (considering P3 as an exogenous variable). The main objective of this system is to predict the pressure at P3 as a function of the pressure at the outlet of valve P2 and the flow rate in the network. This predictive system allows to modulate P2 pressure at the outlet of the valve and detect abnormal water consumption, for example, due to water losses. Additionally, it aims to provide accurate and real-time information on the pressure at P3 to support network maintenance operations, improve leak detection and location, and optimize water use in the distribution network. This predictive tool can be a great ally for water companies that use open control systems, saving costs and improving service for end users.

*2.3. Modelling*

For the analysis of time series, the Box-Jenkins methodology develops statistical models which consider the dependence between data, so that the observation at an instant is modeled as a function of previous values. In this section, the methodology used to select the model that best fits the data series is developed (Figure 8). The Eviews software application, specially designed to work with ARIMA models, has been used for data analysis.

The different variants of ARIMA models are based on the linear dependence of the time series Suhermi et al. [40], Sau et al. [41] as well as on the stationarity of the data used for the Durdu model [42]. Previous studies applied in different field sciences, such as, as energy consumption, stock prices and other economic indicators Sau et al. [41], Durdu [43], Visutsak et al. [44], have shown that ARIMA models can provide accurate results. In the presence of non-linear dependencies of the considered parameters, ARIMA models, by themselves, do not perform adequately. In these cases, the use of artificial neural networks (ANNs) or support vector machines (SVMs), among others, together with ARIMA models, may be more effective Suhermi et al. [40], Aradhye et al. [45], Man-Chun et al. [46]. In view of the technical problem posed, the objective of the present study is to demonstrate the effectiveness of ARIMA models in predicting future pressure values.

The data series available is from February 28, 2021 to August 20, 2022. Thus, the series was divided into two series, the first from February 28, 2021, to March 31, 2022 (38,112 data), was used to generate the model and the second with data from 2022, from April 2022 to August 20, 2022 (13,631 data, was used to validate the model obtained (3.2. Model validation). In both data series, modeling and validation include the summer period, when the city's population increases and also the water requirements. The data are taken by the datalogger every 15 minutes, as explained in section 2.2. Description of the hydraulic control of the water supply in Noja.

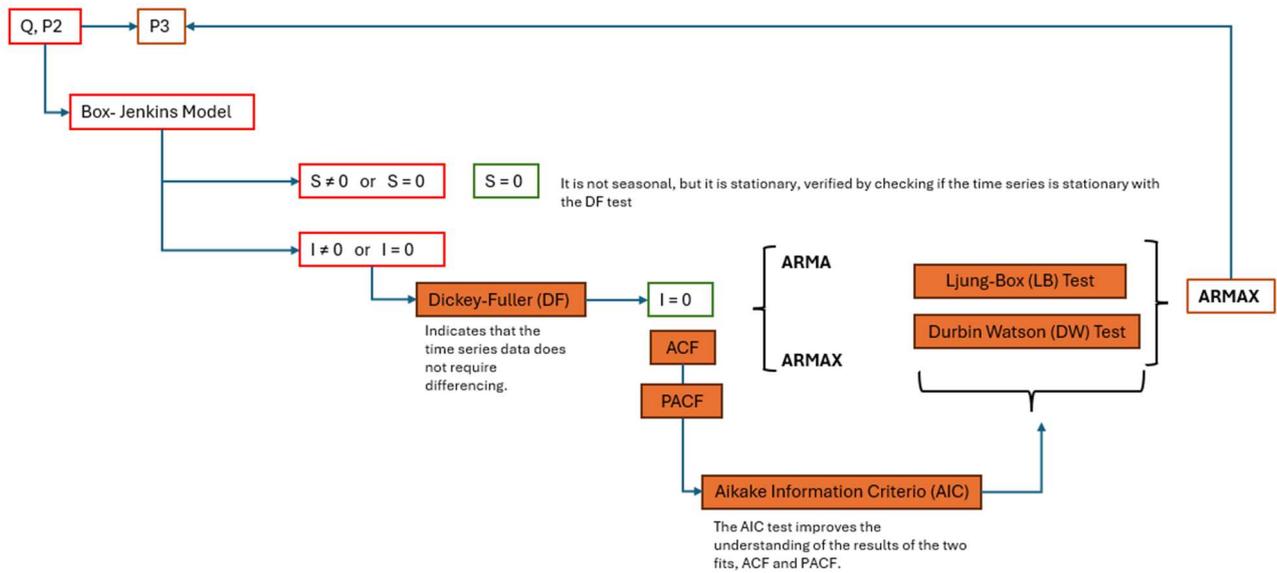

**Figure 8.** Path followed to analyze the data and select the model that best fits the time series used.

The tools to use to select the model are the autocorrelation function (ACF) and the partial autocorrelation function (PACF). With them, the selection of the order of correlation of the autoregressive part of the model (p) is achieved, as well as the degree of relationship in the mean of the errors of the values. Additionally, it can be used to identify periodicities [40]. The delays represented in the graphs (ACF) and (PACF) can take values that can range between ±1, with the delays that present values close to these levels being those that most condition the model.

The augmented Dickey-Fuller (ADF) test is used to determine the stationarity or not of the data set used in the modelling. The null hypothesis of the statistical test assumes the presence of a unit root and therefore the non-stationarity of the values considered. Whether or not the null hypothesis is rejected depends on the significance of the result obtained in the test. If the *p* value is less than the established significance level, which in this study is 0.05 (p<0.05), the rejection of the null hypothesis is confirmed and therefore, the stationarity of the series of values considered is confirmed [48].

To obtain the best model from all possible models, applying the AIC (Akaike Information Criterion) to discern the best model. The selection criterion penalizes complex models, seeking a balance between the fit of the model with the data considered and its complexity. In the comparison of models, the one with the lowest AIC value is chosen.

$$AIC = 2k - 2\ln(\hat{L}) \qquad (1)$$

Where $k$ number of estimated parameters in the model and $\hat{L}$ is the maximum value of the model likelihood function.

Another statistical criterion used in model selection is the Ljung-Box test. This statistical test verifies if there is autocorrelation in the residuals of the adjusted model. The null hypothesis of this statistical test is the non-autocorrelation of the residuals up to a specific lag.

$$Q = n(n+2) \sum_{j=1}^{h} \frac{\rho_j^2}{n-j} \qquad (2)$$

Where $n$ is the sample size, $\rho_j$ the autocorrelation at lag j and h is the total number of lags being tested [49]. In the same sense, another test used to examine the presence of autocorrelation, but first order only, is the Durbin-Watson test [50]. When talking about first-order autocorrelation, it refers to the correlation between adjacent errors. This statistic can take values between 0 and 4. If the statistic takes a value of 2 it means that there is no significant autocorrelation between the first order residuals, especially.

$$DW = \frac{\sum_{t=2}^{n}(e_t - e_{t-1})^2}{\sum_{t=1}^{n} e_t^2} \qquad (3)$$

where the numerator of the statistic $\sum_{t=2}^{n}(e_t - e_{t-1})^2$ is the sum of the squares of the differences between adjacent residues and the denominator, the sum of the squares of all residuals $\sum_{t=1}^{n} e_t^2$.

In the present work, an ARMA and an ARMAX model have been obtained, with two exogenous factors, which are P2, pressure value at the outlet of the reducing valve and flow rate, for the forecast of the pressure at the critical point, called P3.

## 3. Results and Discussion

### 3.1. Model selection.

Before obtaining the models, the time series of the three variables involved were analysed with data from 2021. Figure 9 shows the stationarity analyses necessary for the modelling of the time series. Both the dependent variable (P3) and the exogenous factors (P2, Flow) show the fulfilment of the condition.

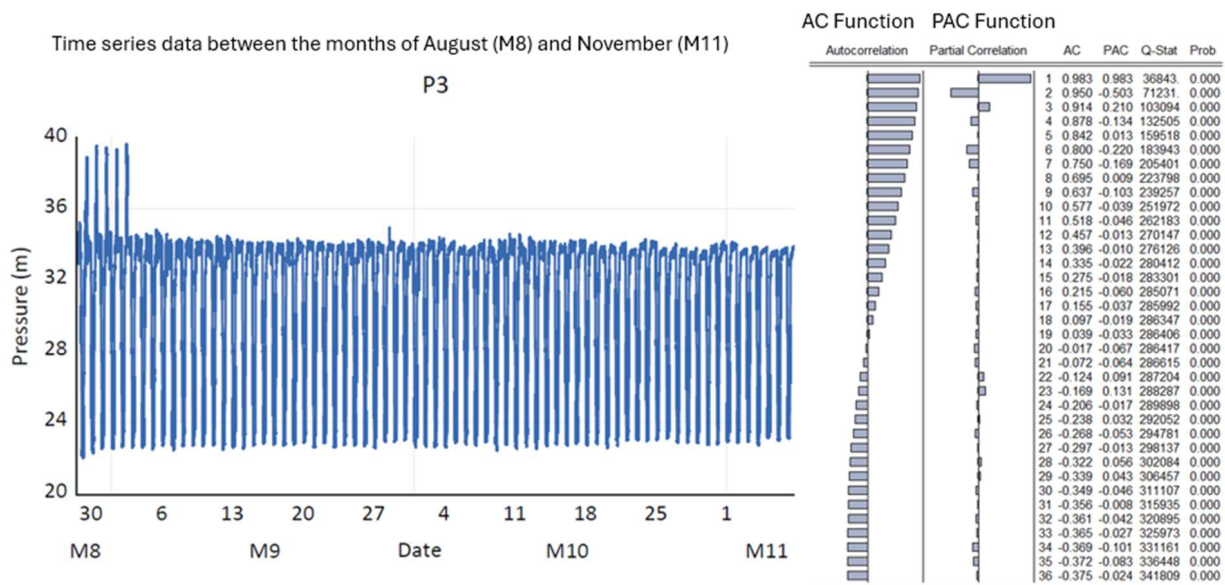

**Figure 9.** Dickey-Fuller Test.

**Figure 10.** Analysis of the P3 series: data sequence, autocorrelations (ACF) and partial autocorrelations (PACF).

Figure 10 shows both part of the plot of the P3 values as a function of time and the autocorrelation and partial autocorrelation plots of the dependent variable P3. From the ACF and PACF plots we identify a model, for the time series of P3, as maximally ARIMA [AR (p), I(d), MA (q)] (4,0,4). To check the adequacy of the model we use the AKAIKE (AIC) criterion [51,52]. In this paper, we will develop ARIMA and ARIMAX models, taking into account the exogenous factors P2 and flow rate. When no differencing is applied, ARIMA models are referred to as ARMA, with the integration (I) equal to zero. The proposed models satisfy this condition.

**ARMAX**

Model Selection Criteria Table
Dependent Variable: P3
Date: 01/21/24   Time: 09:14
Sample: 2/28/2021 00:00 3/31/2022 23:45
Included observations: 38112

| Model | LogL | AIC* | BIC | HQ |
|---|---|---|---|---|
| (4,4)(0,0) | 20193.430... | -1.059059 | -1.056368 | -1.058205 |
| (3,3)(0,0) | 20190.937... | -1.059033 | -1.056790 | -1.058322 |
| (4,3)(0,0) | 20018.094... | -1.049911 | -1.047443 | -1.049128 |
| (3,4)(0,0) | 20018.049... | -1.049908 | -1.047441 | -1.049125 |
| (4,2)(0,0) | 19817.848... | -1.039455 | -1.037212 | -1.038743 |
| (3,2)(0,0) | 19809.924... | -1.039091 | -1.037073 | -1.038451 |
| (2,4)(0,0) | 19806.017... | -1.038834 | -1.036591 | -1.038122 |
| (1,4)(0,0) | 19803.292... | -1.038743 | -1.036725 | -1.038103 |
| (2,3)(0,0) | 19799.944... | -1.038568 | -1.036549 | -1.037927 |
| (1,3)(0,0) | 19781.779... | -1.037667 | -1.035873 | -1.037098 |
| (4,1)(0,0) | 19768.511... | -1.036918 | -1.034899 | -1.036278 |
| (2,2)(0,0) | 19762.358... | -1.036648 | -1.034853 | -1.036078 |
| (3,0)(0,0) | 19756.625... | -1.036399 | -1.034829 | -1.035901 |
| (4,0)(0,0) | 19757.352... | -1.036385 | -1.034591 | -1.035816 |
| (3,1)(0,0) | 19757.004... | -1.036367 | -1.034572 | -1.035797 |
| (1,2)(0,0) | 19736.842... | -1.035361 | -1.033791 | -1.034863 |
| (2,1)(0,0) | 19735.441... | -1.035288 | -1.033718 | -1.034790 |
| (1,1)(0,0) | 19731.196... | -1.035117 | -1.033772 | -1.034690 |
| (2,0)(0,0) | 19644.353... | -1.030560 | -1.029214 | -1.030133 |
| (1,0)(0,0) | 18658.041... | -0.978854 | -0.977733 | -0.978498 |
| (0,4)(0,0) | 15995.010... | -0.838949 | -0.837155 | -0.838380 |
| (0,3)(0,0) | 14159.000... | -0.742653 | -0.741083 | -0.742155 |
| (0,2)(0,0) | 11271.598... | -0.591184 | -0.589838 | -0.590757 |
| (0,1)(0,0) | 6008.166073 | -0.315028 | -0.313906 | -0.314672 |
| (0,0)(0,0) | -6088.549450 | 0.319718 | 0.320615 | 0.320003 |

**ARMA**

Model Selection Criteria Table
Dependent Variable: P3
Date: 01/21/24   Time: 08:59
Sample: 2/28/2021 00:00 3/31/2022 23:45
Included observations: 38112

| Model | LogL | AIC* | BIC | HQ |
|---|---|---|---|---|
| (4,4)(0,0) | -36356.4207 | 1.908397 | 1.910640 | 1.909109 |
| (4,3)(0,0) | -36714.9648 | 1.927160 | 1.929179 | 1.927801 |
| (3,4)(0,0) | -36755.0987 | 1.929266 | 1.931285 | 1.929907 |
| (2,4)(0,0) | -36809.1330 | 1.932049 | 1.933844 | 1.932619 |
| (3,3)(0,0) | -36816.8586 | 1.932455 | 1.934249 | 1.933024 |
| (4,2)(0,0) | -36834.5986 | 1.933386 | 1.935180 | 1.933955 |
| (2,3)(0,0) | -36928.7109 | 1.938272 | 1.939842 | 1.938770 |
| (3,2)(0,0) | -36929.7870 | 1.938328 | 1.939899 | 1.938827 |
| (2,2)(0,0) | -36933.5322 | 1.938473 | 1.939818 | 1.938899 |
| (4,1)(0,0) | -36932.8377 | 1.938489 | 1.940059 | 1.938987 |
| (1,4)(0,0) | -37783.1527 | 1.983110 | 1.984681 | 1.983609 |
| (3,1)(0,0) | -37816.8747 | 1.984828 | 1.986173 | 1.985255 |
| (2,1)(0,0) | -37820.8175 | 1.984982 | 1.986103 | 1.985338 |
| (1,3)(0,0) | -37824.2485 | 1.985215 | 1.986560 | 1.985641 |
| (1,2)(0,0) | -37831.9110 | 1.985564 | 1.986686 | 1.985920 |
| (4,0)(0,0) | -37886.2191 | 1.988467 | 1.989812 | 1.988894 |
| (1,1)(0,0) | -38014.8090 | 1.995110 | 1.996007 | 1.995394 |
| (3,0)(0,0) | -38237.1649 | 2.006831 | 2.007952 | 2.007186 |
| (2,0)(0,0) | -39102.2095 | 2.052173 | 2.053070 | 2.052458 |
| (1,0)(0,0) | -44666.8654 | 2.344137 | 2.344809 | 2.344350 |
| (0,4)(0,0) | -49990.0134 | 2.623636 | 2.624982 | 2.624063 |
| (0,3)(0,0) | -58092.6534 | 3.048785 | 3.049907 | 3.049141 |
| (0,2)(0,0) | -66813.9120 | 3.506398 | 3.507295 | 3.506682 |
| (0,1)(0,0) | -85109.1837 | 4.466424 | 4.467097 | 4.466638 |
| (0,0)(0,0) | -109462.726 | 5.744371 | 5.744819 | 5.744513 |

**Figure 11.** ARMAX Model Selection Criteria Table (left). ARMA Model Selection Criteria (right).

Figure 11 shows, in the left image, the ARMAX models possible from the chosen conditions, and in the right image, the ARMA models. The Eviews 12 software represents the ARMA or ARMAX models with two numbers in brackets. The first bracket shows the degree of the regular (ar) and non-seasonal moving average (ma) part and the second bracket, the degree of the regular (sar) and moving average (sma) part of the seasonal part. Following the AKAIKE criterion, for both the ARMA and ARMAX models, the best model would be (4,4)(0,0). Both without seasonal and non-seasonal differentiation.

For the models with the smallest values when applying the AKAIKE criterion, we check which model meets the Ljung box test. Table 4 shows a summary table of the models with the best AKAIKE criterion and whether they meet Ljung's box test. Table 5 shows the ARMA models with the same information.

**Table 4.** Compliance Ljung-Box ARMAX Model.

| ARMAX Model | LogL | AIC* | BIC | HQ | Ljung-box |
|---|---|---|---|---|---|
| (4,4)(0,0) | 20193.430651 | -1.059059 | -1.056368 | -1.058205 | No |
| (3,3)(0,0) | 20190.937871 | -1.059033 | -1.05679 | -1.058322 | No |
| **(4,3)(0,0)** | **20018.094597** | **-1.049911** | **-1.047443** | **-1.049128** | **Si** |

**Table 5.** Compliance Ljung-Box ARMA Model.

| ARMA Model | LogL | AIC* | BIC | HQ | Ljung-box |
|---|---|---|---|---|---|
| (4,4)(0,0) | -36356.4207 | 1.908397 | 1.910640 | 1.909109 | No |
| (4,3)(0,0) | -36714.9468 | 1.927160 | 1.929179 | 1.927801 | No |
| (3,4)(0,0) | -36755.0987 | 1.929266 | 1.931285 | 1.929907 | No |

The only model that meets the Ljung-box criterion is ARMAX (4,3)(0,0). Figure 12 shows the equation of the model as well as the residual plots. As shown in the left image of Figure 11, the $R^2$ of the model is 0.99888 and the Durbin-Watson statistic, with a value of 1.999668, very close to 2, indicates no autocorrelation between contiguous residuals. The image on the right shows compliance with the Ljung-box criterion up to lag 9.

```
Dependent Variable: P3
Method: ARMA Maximum Likelihood (BFGS)
Date: 01/21/24   Time: 18:22
Sample: 2/28/2021 00:00 3/31/2022 23:45
Included observations: 38112
Convergence achieved after 72 iterations
Coefficient covariance computed using outer product of gradients

Variable      Coefficient   Std. Error   t-Statistic   Prob.
C              6.202652     0.021390     289.9810      0.0000
FLOW          -0.059023     0.000153    -384.7124      0.0000
P2             1.020825     0.000702     1454.768      0.0000
AR(1)          2.847282     0.007309     389.5612      0.0000
AR(2)         -2.903856     0.017676    -164.2810      0.0000
AR(3)          1.178743     0.016322     72.21849      0.0000
AR(4)         -0.124493     0.005890    -21.13676      0.0000
MA(1)         -2.222669     0.007183    -309.4283      0.0000
MA(2)          1.709916     0.012190     140.2668      0.0000
MA(3)         -0.454516     0.006554    -69.34734      0.0000
SIGMASQ        0.020478     3.27E-05     625.5269      0.0000

R-squared              0.998880   Mean dependent var     30.35965
Adjusted R-squared     0.998880   S.D. dependent var      4.276810
S.E. of regression     0.143122   Akaike info criterion  -1.049911
Sum squared resid      780.4538   Schwarz criterion      -1.047443
Log likelihood         20018.09   Hannan-Quinn criter.   -1.049128
F-statistic            3399327.   Durbin-Watson stat      1.999668
Prob(F-statistic)      0.000000

Inverted AR Roots    .97    .86-.27i    .86+.27i    .16
Inverted MA Roots    .76    .73-.25i    .73+.25i

Date: 01/21/24   Time: 09:32
Sample (adjusted): 2/28/2021 00:00 3/31/2022 23:45
Q-statistic probabilities adjusted for 7 ARMA terms

Autocorrelation   Partial Correlation       AC      PAC    Q-Stat   Prob*
                                      1   0.000   0.000    0.0008
                                      2  -0.001  -0.001    0.0737
                                      3   0.004   0.004    0.6593
                                      4   0.000   0.000    0.6684
                                      5  -0.006  -0.006    2.0133
                                      6   0.005   0.005    2.9040
                                      7  -0.004  -0.005    3.6737
                                      8  -0.001  -0.001    3.7294   0.053
                                      9  -0.005  -0.005    4.8202   0.090
                                     10   0.014   0.014   12.103    0.007
                                     11   0.007   0.007   14.193    0.007
                                     12  -0.002  -0.002   14.388    0.013
```

**Figure 12.** Model equation. Test Ljung-Box.

*3.2. Model validation.*

Once the model that best fits the type of time series analyzed is selected, it is validated with real data. The idea is to analyze the forecasted values (P3 (P3F)) with the model, compare them with the actual values recorded by the data logger in the field, and use the ANOVA method to determine if there are significant differences between the data.

Table 6 shows the normalised analysis of variance between the values predicted by the model and the actual values. With a significance level of 95% the statistical test result shows that there is no difference between the predicted values and the actual values.

**Table 6.** One-factor Normalized Analysis of Variance (ANOVA).

| Origin of variations | Sum of squares | Degrees of freedom | mean squares | F | Probability | Critical value for F |
|---|---|---|---|---|---|---|
| between groups | 46.52 | 1 | 46.52 | 2.13 | 0.14 | 3.84 |
| within the groups | 2256212.31 | 103484 | 21.80 | | | |
| Total | 2256258.84 | 103485 | | | | |

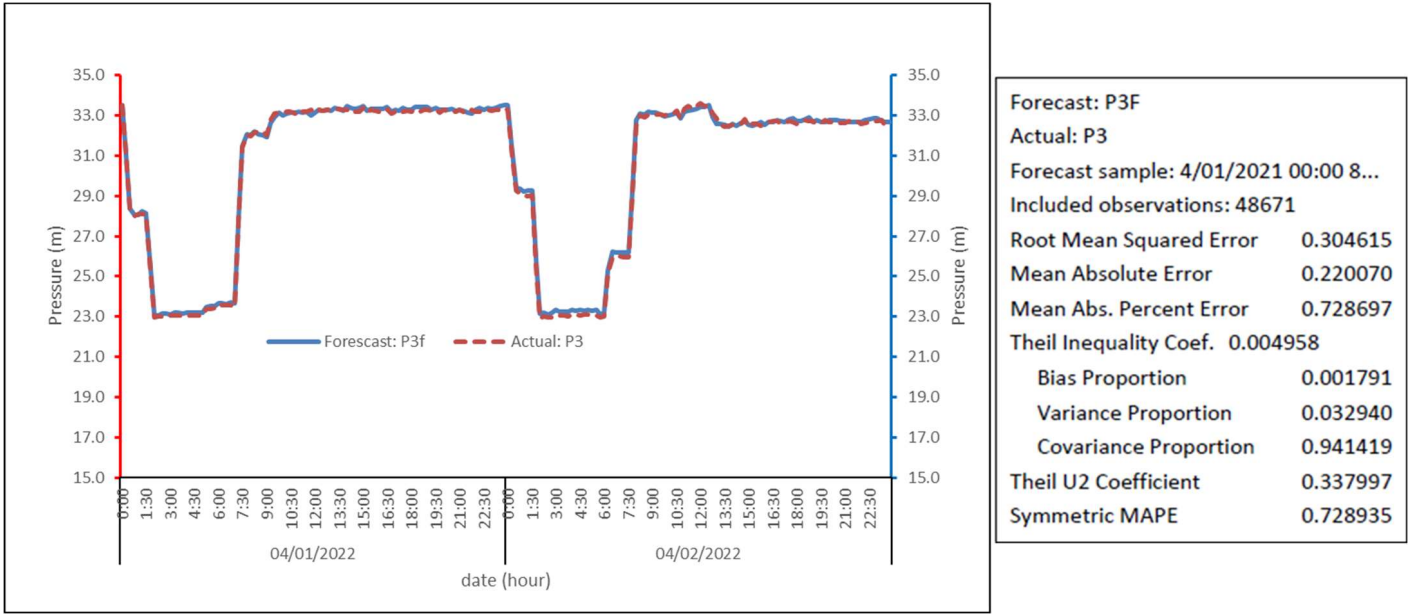

**Figure 13.** Graph forecast P3 (P3F) and actual P3.

To analyze the accuracy of the predictive model, as well as its comparison with other models, the root mean squared error (RMSO), Mean Absolute Error (MAE), Mean Absolute Percent Error (MAPE) and Theil Inecuality Coefficient (U) statistics are considered:

$$RMSO = \sqrt{\frac{\sum_{i=1}^{n}(Actual - Forecast)^2}{n}} \qquad (4)$$

$$MAE = \frac{\sum_{i=1}^{n}|Forecast - Actual|}{n} \qquad (5)$$

$$MAPE = \frac{1}{N} \cdot \sum_{i=1}^{n}\left|\frac{Actual - Forecast}{Actual}\right| \cdot 100 \qquad (6)$$

$$U = \frac{\sqrt{\frac{1}{N} \cdot \sum_{i=1}^{n}(Forecast - Actual)^2}}{\sqrt{\frac{1}{N} \cdot \sum_{i=1}^{n}(Forecast)^2} + \sqrt{\frac{1}{N} \cdot \sum_{i=1}^{n}(Actual)^2}} \qquad (7)$$

Figure 13 shows the representation of part of the values of the real P3 data series, measured in the installation, and the P3 values predicted by means of the ARMAX model obtained. To represent the real P3 data series, a dashed red line has been chosen, superimposed on the forecast series, which is blue and continuous, with no differences. In addition to the ANOVA analysis carried out on the series of forecast data and real data, figure 13 shows both the mean square error of the series compared, giving this statistic a value of 0.3, and the mean absolute error, with a value of 0.22, both very low values, indicating that the difference between the real values and the forecast values is very low.

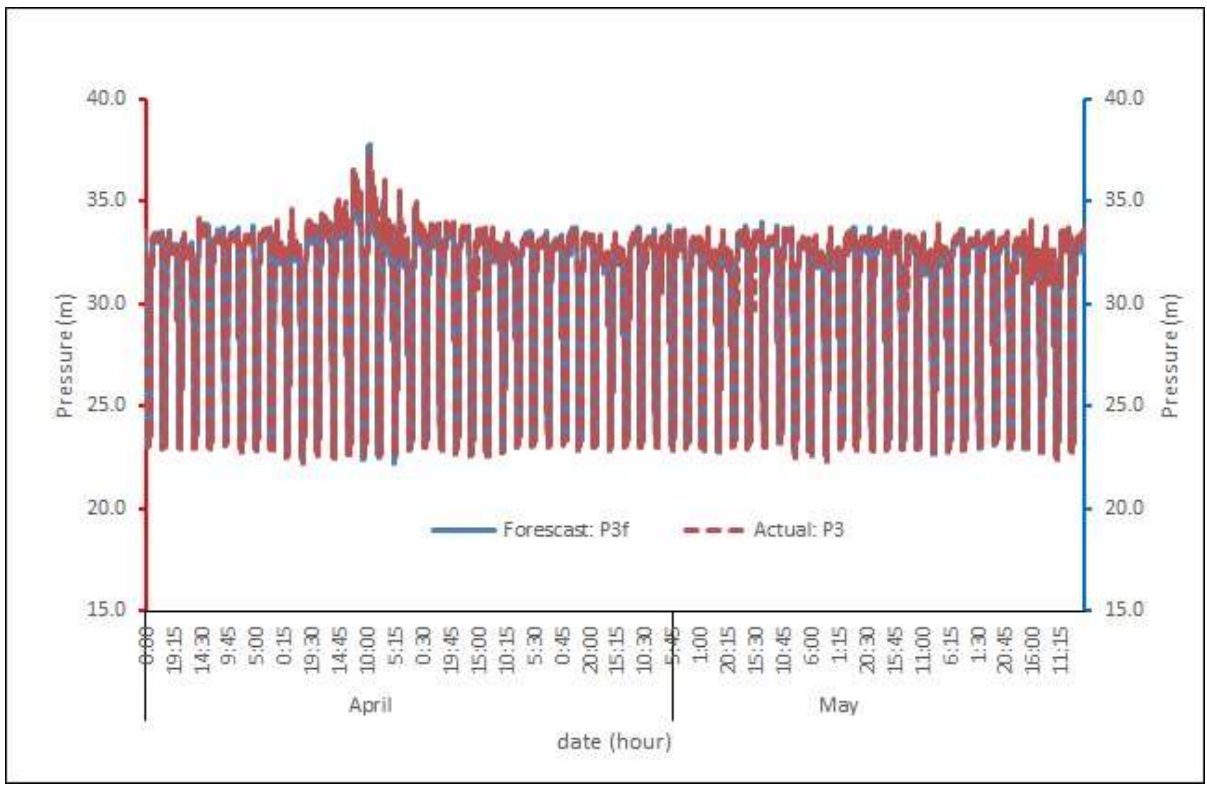

**Figure 14.** Graph forecast P3 (P3F) and actual P3 during April and May 2022.

Other results that confirm the good fit of the model and its forecasts include a MAPE of 0.72% and a low Theil Inequality Coefficient (0.004958), indicating strong model accuracy. The Theil Inequality Coefficient can be decomposed into three proportions that explain the source of error: bias, variance, and covariance [53,54]. The closer the bias and variance proportions are to zero, the less the error is due to bias or variance between the predicted and actual values; in this study, these proportions are 0.002 and 0.033, respectively. Conversely, the closer the covariance proportion is to 1, the better correlated the predictions and actual values are. In the ARMAX model, it is near unity, with a value of 0.94 (Figure 13). Figure 14 illustrates a two-month period of predicted and actual values measured at the facility, although the forecast extends from April to August 20, 2022 (covering 13,631 data points). Data contained in the dataset analyzed using the ANOVA analysis is shown in Table 7.

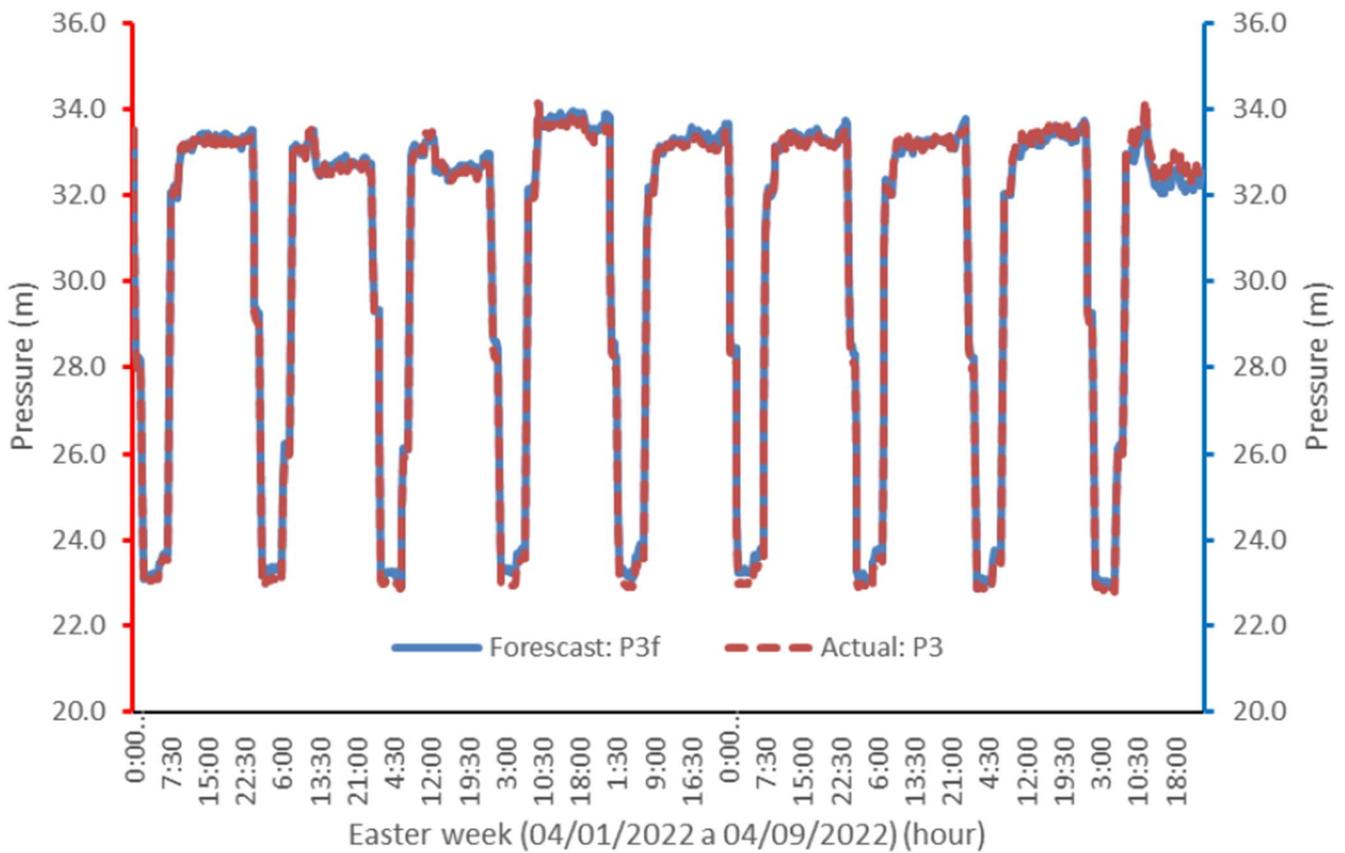

**Figure 15.** Graph forecast P3 (P3F) and actual P3 during Easter week 2022.

Figure 15 shows the predicted values for P3 and the actual measured values for P3. In this case, the predicted time period corresponds to the Easter week of the year under study. In the year 2022 it coincided in the period between 04/01/2022 and 04/09/2022 and that time of the year, the population of Noja experiences a substantial increase in population. Figure 15 shows the adjustment of the predicted values with the real values.

**Table 7.** One-factor Normalized Analysis of Variance (ANOVA).

| Origin of variations | Sum of squares | Degrees of freedom | mean squares | F | Probability | Critical value for F |
|---|---|---|---|---|---|---|
| between groups | 1.97 | 1.00 | 1.97 | 0.12 | 0.73 | 3.85 |
| within the groups | 28650.52 | 1726.00 | 16.60 | | | |
| Total | 28652.50 | 1727.00 | | | | |

Table 7 shows the anova study carried out on the data for the Easter week period. At a significance level of 95%, the probability value shows a value equal to 0.73 and the null hypothesis of the study cannot be rejected, so there are no differences between the mean values of the predicted and actual P3 data series.

The P3 critical point pressure information provided by the ARMAX model can be extremely useful for the management of water distribution networks. Unexpected variations in pressure at the P3 critical point can indicate the presence of leaks, abnormal consupmtion or other anomalies in the network. Furthermore, the implementation of a continuous feedback system has the potential to substantially augment the predictive capacity and adaptability of the ARMAX model. By periodically updating the model with real-time or scheduled data from the network, the system can recalibrate its parameters in real time, thereby ensuring its reliability and accuracy when facing seasonal variations or unanticipated changes in operational conditions. This approach is consistent with best practices in resource management, facilitating the early detection of leaks and the optimisation of pressure control strategies over time. By continuously monitoring the pressure at P3 using the ARMAX model, operators can detect such variations early and take preventive action to repair leaks or other problems before they develop into more serious situations. Another important aspect is that maintaining the pressure in the network within optimal ranges is crucial to ensure efficiente and reliable supply of drinking water. Using the pressure forecasts in P3 provided by the ARMAX model, operators can adjust the settings of pressure control valves and other regulators to optimise pressure throughout the network. This not only helps reduce the risk of leaks and infrastructure damage due to excessive pressure, but also improves the overall operational efficiency of the system. By analysing historical and real-time P3 pressure data provided by the ARMAX model, operators can identify trends and patterns of behaviour that indicate the need for preventive maintenance in specific areas of the network. This allows them to schedule maintenance more effectively and proactively, reducing the risk of unexpected failures and minimising system downtime.

The successful implementation of the ARMAX model is contingent upon the availability of high-quality data. Inconsistent, incomplete, or noisy datasets can introduce significant challenges, affecting the model's ability to make accurate predictions. Ensuring data quality necessitates the establishment of robust data collection systems and the regular maintenance of sensors and logging equipment.

Furthermore, adapting the model to different contexts poses additional challenges. Variations in network configurations, demand patterns, and operational conditions may require recalibration or the inclusion of additional exogenous factors to maintain the model's accuracy. This underscores the necessity for future research to investigate the model's performance across a range of water distribution systems and operational scenarios, as well as the potential for hybrid approaches combining predictive analytics with real-time feedback mechanisms.

Applying this methodology in other regions or countries presents additional challenges, particularly in systems with different infrastructure characteristics or operational constraints. For example, networks in regions with limited data availability or less advanced monitoring systems may struggle to achieve the data quality required for model training and validation. In addition, variations in topography, climate and population dynamics may require model adaptations to ensure accurate predictions and robust performance in different contexts. These considerations highlight the need for further research to assess the generalisability and scalability of the model in different scenarios.

Finally, it is important to highlight that the proposed ARMAX model significantly contributes to sustainability in water distribution systems. By optimizing pressure management and improving leak detection, it reduces water losses, conserving this critical resource and decreasing the energy demand associated with water treatment and pumping. This aligns with global efforts to promote sustainable water management practices, particularly in areas facing increasing water scarcity.

Additionally, these improvements have a direct positive impact on local communities. A more efficient water distribution system ensures a reliable supply of clean water, even during periods of peak demand, such as those observed in Noja during the summer season. This enhances the quality of life for residents and reduces the risks associated with intermittent water supplies, such as health concerns or disruptions to daily activities. Furthermore, by lowering operational costs through reduced water loss and maintenance needs, the proposed approach can enable water utilities to allocate resources to other critical infrastructure improvements, further supporting community well-being.

*3.3. Model application*

It must be highlithed that the model can apply only in areas fed through one single point. For mulifeed areas, with more than one entry of water, this model would require to be reevaluted and adapted. However, the most common application in water distribution system are single fed areas.

Also, the model would not apply in areas which are not well isolated or where the boundaries are changing continuosly. This would involve major changes in the flow and thus the headloss between the outlet of the pressure reducing valve and the critica point of the area.

## 4. Conclusions

One of the main issues in potable water distribution networks is leak management. A thorough understanding of the network and effective sectorization are fundamental tools for companies in the sector to manage their assets properly. In recent years, water management companies have been making significant efforts to reduce leakage rates and ensure long-term supply to the population due to the challenges posed by climate change, population growth, and water scarcity experienced in recent years.

Leak management has a dual perspective: on the one hand, the economic loss that these unbilled water losses represent for public, private, or mixed companies. On the other hand, it represents a waste of a non-renewable resource, especially during drought periods as experienced in recent years in Spain due to climate change. There are various tools to detect leaks in distribution networks, and as seen in this study, pressure management is a very interesting parameter for predicting potential network leaks. Such tools are useful in systems like the one analyzed in Noja, using open-loop control, where the use of predictive models generates considerable interest among network management companies.

As analyzed in the study, pressure management, achieved through pressure reducing valves (PRVs), plays a crucial role in reducing and preventing water loss in distribution piping systems. By controlling pressure levels, leakage rates can be effectively managed, ensuring optimal water supply and distribution. Critical point pressure control systems, particularly those employing open-loop approaches, offer sophisticated solutions for maintaining optimal pressure levels at critical network points.

Different models based on the Box & Jenkins methodology and its variants have been analyzed. This article proposes a new open-loop critical point system in which ARMA and ARMAX models are constructed and validated to overcome the difficulties of having real-time pressure data at the system's critical point. The best model is generated using the pressure at the PRV outlet and the flow signal as exogenous variables, and the pressure level at the critical point as the dependent variable. The attempt to model the pressure data series at the critical point without considering exogenous factors was rejected as it did not meet the Ljung-Box criterion. Among all the selected ARMAX models, those with the highest value for the AKAIKE criterion, only one model met the Ljung-Box criterion up to lag number 9. The model shows a good $R^2$ value of 0.99888, indicating the best fit, even during the holiday season from June to September, a period with a significant increase in demand due to the increased population in the supplied area. Another time of great demand for water consumption, due to the growth of the population in the area of Noja, is Easter week. As shown in Figure 15 and Table 7, with the ANOVA analysis, the prediction made with the ARMAX model is indistinguishable from the evolution followed by the real values.

As a future improvement, incorporating a continuous feedback system to update the ARMAX model with real-time data would ensure its long-term effectiveness and adaptability. Such a system would not only support dynamic adjustments to operational changes but also enhance proactive management of anomalies, making water distribution networks more sustainable and resilient.

Another point of future research would involve creating a model with similar accuracy requiring less amount ot datapoints entry to build the algorithm.

In this work, only the ARIMA model has been considered, as well as its variant ARMAX, the new trends of implanting mixed models, between the model studied and artificial neural networks (ANNs), is a clear option for study, although the results obtained with the use of the ARMAX model are so good. Future studies could explore the integration of mixed models with artificial neural networks (ANNs), leveraging boundary conditions to better understand their influence on pressure variations like those observed at P3. This approach could improve anomaly classification and enhance predictive capabilities for pressure management in complex water distribution systems.

Finally, it is important to highlight the limitations and future opportunities of this line of research. The main limitation is the type and size of the network with control at a single PRV. It would be interesting to apply this model to larger networks with several control points to examine the robustness of predictive models based on the analysis of data series. While the study focuses on a single PRV in the Noja network, this limitation highlights the challenges of scaling the approach to larger, more complex networks. Larger networks often have multiple critical points, interacting PRVs, and varying hydraulic conditions, which may require more sophisticated models or hybrid approaches combining real-time data with predictive analytics. Additionally, the dependency on historical data for model training could pose challenges in networks with incomplete or inconsistent datasets, limiting the robustness of predictions in such scenarios. Future research should explore how the ARMAX model performs in multi-PRV systems or networks with higher levels of variability, potentially integrating additional exogenous factors or advanced feedback systems to improve adaptability and scalability.

This model is ready to be implemented in Noja's pressure control system. The advantages of applying the ARMAX model in critical point pressure management are clear, including improved leak detection and location, enhanced network maintenance operations, and optimized water use efficiency. The implementation of the ARMAX model in the Noja water distribution network would have significant practical impacts. From an economic perspective, the ARMAX model has the potential to reduce water loss and associated costs by facilitating more efficient and precise leak detection. From an environmental perspective, the model fosters sustainable water management by optimising pressure levels and reducing unnecessary water wastage, thereby contributing to the conservation of this vital resource. Operationally, the enhanced predictive capabilities would support more efficient network maintenance, improving overall reliability and service quality for end users. The model's potential to address key challenges in water distribution is evident in these impacts, and it provides a scalable and sustainable solution for similar systems worldwide. By accurately predicting pressure levels at critical points, operators can proactively address issues such as leaks, abnormal consumption, and network anomalies. The ARMAX model provides valuable insights for water distribution network operators, enabling them to detect pressure variations early and take preventive measures to mitigate potential problems. Furthermore, by optimizing pressure levels throughout the network, operators can reduce the risk of infrastructure damage, improve operational efficiency, and minimize system downtime. By leveraging advanced technologies and data-driven solutions, water utilities can enhance their capacity to ensure a reliable water supply, minimize water loss, and promote the sustainability of water resources for future generations. This research will enhance the reliability of predictions, enabling management company planners to make more robust decisions.

**Acknowledgments:** The authors would like to thank GS Inima water utility for their contribution, availability, criticism, and support to materialize this paper.